\documentstyle[12pt]{article}
\newcommand{\be}{\begin{equation}}
\newcommand{\ee}{\end{equation}}
\newcommand{\bea}{\begin{eqnarray}}
\newcommand{\eea}{\end{eqnarray}}
\newcommand{\ba}{\begin{array}}
\newcommand{\ea}{\end{array}}
\newcommand{\bmat}{\left(\ba}
\newcommand{\emat}{\ea\right)}

\newcommand{\pole}{\frac{1}{\varepsilon}\frac{g_B^2}{16\pi^2}}
\newcommand{\alpole}{\frac{\widehat\alpha}{\varepsilon}}
\newcommand{\gb}{g_B^{}}
\newcommand{\gr}{g_R^{}}
\newcommand{\norsl}{\normalsize\sl}
\newcommand{\norsc}{\normalsize\sc}
\begin{document}
%-------------------- Title page ----------------------------------
\begin{titlepage}

\title{QCD Analysis of Twist-4 Contributions \\
       to the $g_1$ Structure Functions }

\author{
\norsc  Hiroyuki KAWAMURA \\
\norsl  Dept. of Fundamental Sciences\\
\norsl  FIHS, Kyoto University\\
\norsl  Kyoto 606-01, JAPAN\\
\\
}

\date{}

\maketitle
\vspace{2cm}
\begin{abstract}
{\normalsize
We analyze the twist-4 contributions to Bjorken and Ellis-Jaffe 
sum rules for spin-dependent structure function $g_1(x, Q^2 )$. 
We investigate the anomalous dimensions of the twist-4 operators 
which determine the logarithmic correction to the $1/Q^2$ behavior 
of the twist-4 contribution by evaluating off-shell Green's functions 
in both flavor non-singlet and singlet case. It is shown that the 
EOM operators play an important role to extract the anomalous 
dimensions of physical operators.
The calculations to solve the operator mixing of higher-twist 
operators are given in detail. }
\end{abstract}

\begin{picture}(5,2)(-310,-615)
\put(15,-125){KUCP-96}
\put(15,-140){June, 1996}
\end{picture}

\vspace{2cm}
\leftline{\hspace{1cm}hep-ph/9605466}
 
\thispagestyle{empty}
\end{titlepage}
\setcounter{page}{1}
\baselineskip 24pt
%----------------------- Text -----------------------------------

\section{Introduction}
The polarized structure functions $g_1$ and $g_2$ for nucleons are 
measured by recent experiments at CERN \cite{EMC,SMC} and 
SLAC \cite{E142,E143}. These functions provide us 
the non-trivial spin structures of nucleons.  
Especially $g_1$ function has direct partonic interpretations, 
and it turned out that very relativistic pictures hold for nucleon 
spin \cite{REV}.
In the framework of the operator product expansion and the renormalization
group method based on QCD, we can derive sum rules for n-th moments 
of $g_1^{p,n}(x,Q^2)$, where the first moments of $g_1^{p,n}(x,Q^2)$ are 
given by the Ellis-Jaffe sum rules \cite{EJ} and that of $g_1^{p-n}(x,Q^2)$ 
leads to the Bjorken sum rule \cite{Bj}.  
In the deep inelastic scattering, the perturbative QCD has been tested so far
for the effects of the leading twist operators, namely twist-2 operators, for 
which the QCD parton picture holds. Now, the twist-4 operators give
rise to $O(1/Q^2)$ corrections to the first moment of $g_1(x,Q^2)$, 
which may have some contribution in the region of $Q^2$ of the recent 
experiments. Their contributions correspond to the correlations 
between quarks and gluons.

In this paper we investigate the renormalization of the twist-4
operators, which are relevant for
the first moment of the nucleon spin structure functions 
$g_1(x,Q^2)$, 
and obtain their anomalous dimensions. From this calculations 
we determine the logarithmic 
correction to the $1/Q^2$ behavior of the twist-4 operator's
contribution in the first moment of $g_1(x,Q^2)$.
We study the renormalization of the composite operators 
by evaluating Green's functions taking the external lines off-shell 
so that we can avoid the subtle problems of IR divergence.      
The general feature in renormalization of higher twist operators 
by off-shell Green's functions is that there occurs the mixing 
among the physical operators and the EOM operators 
which are proportional to the equation of motion and 
what we call the `BRS-exact' operators which contain ghost 
operators \cite{JL,COLL}. Although the physical matrix elements 
of EOM and BRS-exact operators vanish, we need to consider properly 
their contributions to the counterterms to extract the anomalous 
dimensions of the physical operators. 
As we show later, the tree vertices 
of these unphysical operators may have components of the same tensor 
structures as that of the physical one. So we need to identify 
correctly the operator basis and separate the divergence of 
the radiative corrections into two parts  
that correspond to the physical operators and the unphysical 
operators, respectively.
As for the operator mixing, some formal arguments can be given in 
the gauge theory \cite{JL,COLL}, and in this paper we confirm these 
theorems explicitly.

In the following sections, we obtain all mixing matrix elements for 
flavor non-singlet part, and only the physical 
one for flavor singlet part. In both cases, it turns out that there
exists only one physical operator \cite{SV}. 

\section{Twist-4 contribution to the first moment of $g_1(x,Q^2)$}
The polarized deep inelastic scattering is described by the antisymmetric 
part of the hadronic structure tensor $W_{\mu\nu}^A$ 
given in terms of two spin structure functions $g_1(x,Q^2)$ 
and $g_2(x,Q^2)$ as
\be
W^A_{\mu\nu} = \varepsilon _{\mu\nu\lambda\sigma} q^{\lambda}
                \left\{ s^{\sigma} \frac{1}{p\cdot q} g_1(x, Q^2 )
                + ( p\cdot q s^{\sigma} - q\cdot s p^{\sigma} )
                 \frac{1}{(p\cdot q)^2} g_2 (x, Q^2 )\right\},
\ee
where
$q$ is the virtual photon momentum and $p$ is the nucleon momentum.
$x$ is the Bjorken variable $x=Q^2/2p\cdot q=Q^2/2M\nu$ and 
$q^2=-Q^2$. $M$ is the nucleon mass and
$s^{\mu}={\overline u}(p,s)\gamma^{\mu}\gamma_5u(p,s)$ is the covariant
spin vector of the nucleon.

Now the first moment of the $g_1(x,Q^2)$ structure functions for proton and
neutron turns out to be up to the power correction of order $1/Q^2$:
\bea
&&\hspace{-0.8cm}\Gamma_1^{p,n}(Q^2)\equiv \int_0^1 g_1^{p,n}(x,Q^2)dx 
\nonumber\\
&&\hspace{-0.8cm}=(\pm\frac{1}{12} g_A+\frac{1}{36}a_8)
\bigl ( 1-\frac{\alpha_s}{\pi}+ {\cal O}(\alpha_s^2)\bigr )
+\frac{1}{9}\Delta\Sigma\ \bigl ( 
1-\frac{33-8N_f}{33-2N_f}\frac{\alpha_s}{\pi}+{\cal O}(\alpha_s^2)
\bigr ),                                         
\label{gam1}
\eea
where $g_1^{p(n)}(x,Q^2)$ is the spin structure function of the 
proton (neutron) and the plus (minus) sign is for proton (neutron).
On the right-hand side, $a_3=g_A \equiv G_A/G_V$ is the ratio of 
the axial-vector to vector coupling constants. 
Here we assume that the number of active 
flavors in the current $Q^2$ region is $N_f=3$. Denoting
$\langle p,s|\overline{\psi}\gamma_\mu\gamma_5\psi|p,s\rangle
=\Delta q s_\mu$, the flavor-$SU(3)$ octet and singlet part, $a_8$ and 
$a_0=\Delta\Sigma$ are given by
\be
a_8 \equiv \Delta u +\Delta d -2\Delta s, \qquad
\Delta\Sigma \equiv \Delta u +\Delta d +\Delta s . \nonumber 
\ee
The scale-dependent density $\Delta\Sigma$ evolves as 
\be
\Delta\Sigma(Q^2)=\Delta\Sigma(\infty)\left(1+\frac{6N_f}{33-2N_f}  
\frac{\alpha_s(Q^2)}{\pi}\right).
\ee

Taking the difference between $\Gamma_1^p$ and $\Gamma_1^n$ leads to the QCD
Bjorken sum rule: 
\be 
\Gamma_1^{p-n} \equiv 
\int_0^1 dx \bigl [ g_1^p(x,Q^2)-g_1^n(x,Q^2) \bigr ]
=\frac{1}{6} g_A \bigl [ 1-\frac{\alpha_s(Q^2)}{\pi}+
{\cal O}(\alpha_s^2) \bigr ].
\label{bj}
\ee
The first order QCD correction was calculated in \cite{KOD,K} and 
the higher order corrections were given in \cite{GL}. 

Now, the twist-4 operator gives rise to $O(1/Q^2)$ corrections to
the RHS of $(\ref{gam1})$. Their flavor decompositions are 
just the same as those of leading twists. From renormalization 
group analysis, their $\log{Q^2}$ dependences take the following form 
in leading-log approximation.
\be
\delta\Gamma_{1,tw-4}^{p,n}(Q^2)= -\frac{8}{9Q^2}\Bigl [ \{ \pm\frac{1}{12} f_3+
\frac{1}{36}f_8\}
\left(\frac{\alpha_s(Q_0^2)}{\alpha_s(Q^2)}\right)^
{-\gamma_{NS}^0/2\beta_0}\hspace{-0.1cm}+\frac{1}{9}f_0
\left(\frac{\alpha_s(Q_0^2)}{\alpha_s(Q^2)}\right)^
{-\gamma_{S}^0/2\beta_0}\Bigr ], 
\label{tw4}
\ee
where $f_0$, $f_3$ and $f_8$ are the twist-4 counter parts of $a_0$, $a_3$ 
and $a_8$. 
\be
\langle p,s |R_{\sigma}^i|p,s\rangle = f^i s_{\sigma}
\ee
$f_i$'s are scale dependent and here they are those at
$Q_0^2$. Here we assume that there exists only one physical
operator for each flavor as we show later. $\gamma_{NS}^0$ and 
$\gamma_{S}^0$ are the coefficients of the one-loop anomalous
dimensions for flavor non-singlet and singlet operators respectively.
They are obtained from the renormalization constants of the 
corresponding operators.    
The magnitude of these corrections depends on the reduced matrix elements
$f_i$ which are not calculable by perturbative QCD. 

On the other hand, target mass effects also give rise to power 
corrections to these sum rules. They can be estimated in full 
order of $M^2/Q^2$ 
by taking the difference between the Nachtmann moments 
and the usual 1st moments \cite{HKU}.   
\bea
&&\Delta\Gamma_1 \equiv\Gamma_1(M^2\neq 0) - \Gamma_1(M^2= 0) \nonumber\\
&&= \int_0^1dx \bigl \{\frac{5}{9}\frac{\xi^2}{x^2}+
\frac{4}{9}\frac{\xi^2}{x^2}\sqrt{1+\frac{4M^2x^2}{Q^2}}-1\bigr\}
g_1(x,Q^2) 
-{4\over 3}\int_0^1 dx {{\xi^2}\over{x^2}}{{M^2x^2}\over{Q^2}}
g_2(x,Q^2).
\eea
It should be noted that we need also $g_2(x,Q^2)$ to estimate
$\Delta\Gamma_1$, and this is because we could not obtain their values  
until the recent experiments of $g_2(x,Q^2)$ came out.
From these experiments \cite{SMC,E143} we get, 
\be
\Delta\Gamma^p_1 =  -0.001\pm 0.002  ,\hspace{1.0cm} \Delta\Gamma^n_1 = -0.001\pm 0.002 \hspace{1cm}
(Q^2=2.5\mbox{GeV}^2).
\ee
So we can conclude that the target mass effects are negligible 
for the Bjorken sum rule and Ellis-Jaffe sum rule  
even in this lower $Q^2$ region because they amount to less than 
a few percent of $\Gamma_1$. 

In the following sections we calculate the anomalous dimension of 
the higher twist operators which determine the logarithmic $Q^2$ dependence
of the twist-4 terms in the first moments of $g_1(x,Q^2)$. 

%section 3

\section{Flavor non-singlet part}
We now consider the renormalization of the operators.
Let us consider the off-shell Green's function of twist-4 composite operators
keeping the EOM operators as independent operators.
Thus we can avoid the subtle infrared divergence which may appear in the
on-shell amplitude with massless particle in the external lines.

Another advantage to study the off-shell Green's function is that
we can keep the information on the operator mixing problem.
And further, the calculation is much more straightforward than the one 
using the on-shell conditions.

From general arguments it is known that there appear three types of operators 
which participate in renormalization of gauge invariant operators. 
\cite{JL,COLL,H}.\\ 

(1) Gauge invariant operators $R_i$ which appear in the operator
product expansion. \\
\hspace*{1.3cm}We call them physical operators for they have 
non-zero physical matrix \\ \hspace*{1.3cm}elements.
 
(2) EOM operators $E_i$ whose physical matrix elements vanish.

(3) BRS-exact operators $A_i$ whose physical matrix elements also vanish.\\

These operators mix with each other through renormalization, and thief renormalization
matrix is to be triangular.
\be
\bmat{c} 
R_i \\ E_i \\ A_i
\emat_{\hspace{-0.1cm}R}
=
\bmat{cccc}
Z_{RR} & Z_{RE} & Z_{RA} \\
 0     & Z_{EE} & Z_{EA} \\
 0     & 0      & Z_{AA} \\
\emat
\bmat{c} 
R_i \\ E_i \\ A_i  
\emat_{\hspace{-0.1cm}B}.
\ee
We should note that only $Z_{RR}$ has physical meaning among 
these renormalization matrix elements because of the triangularity. 

Now for renormalization of the twist-4 operators, at first we need  
to identify the physical operators and other operators which mix with them.  
As can be seen from the dimensional counting, there is no contribution from 
the four-fermi type twist-4 operator to the first moment of $g_1(x,Q^2)$.
The only relevant twist-4 operators turn out to be 
of the form bilinear in quark fields and linear in the gluon field strength.
This is in contrast to the unpolarized case, where both types
of twist-4 operators contribute. 

Operators which mix with each other need to have same properties 
such as dimension, spin and other quantum numbers. 
The relevant operators in our case has the following properties;
It is dimension 5 and spin 1. Its parity is odd and it has to
satisfy the charge conjugation invariance.
The flavor non-singlet operators are bi-linear in fermion fields.
We consider the gauge variant EOM operators as well, but BRS-exact 
operators don't mix because of flavor symmetry.  

The parity and charge conjugation condition requires the relevant operators 
to have an odd number of gamma matrices together with $\gamma_5$ or 
$\varepsilon_{\mu\nu\rho\sigma}$.
The possible twist-4 operators bilinear in $\psi$ and ${\overline\psi}$
are of the form, $O_{\sigma}=\overline{\psi}M_{\sigma}\psi$ where
dim$O_\sigma=5$, namely dim$M_\sigma=2$, and we have 
$M_\sigma=\gamma_5\gamma_\sigma D^2, 
g\widetilde{G}_{\sigma\mu}\gamma^{\mu}, \gamma_5(D_\sigma\not{\!\!D}
+\not{\!\!D}D_\sigma), \gamma_5(\partial_\sigma\not{\!\!D}
+\not{\!\!D}\partial_\sigma)$.
Hence we have the following operators which satisfy the above conditions:
\bea
  R_1^\sigma &=& -\overline{\psi}\gamma_5
       \gamma^{\sigma}D^2\psi  ,                \nonumber\\
  R_2^\sigma &=& g \overline{\psi}
       \widetilde{G}^{\sigma\mu}\gamma_{\mu}\psi ,           \nonumber\\
  E_1^\sigma &=& \overline{\psi} \gamma_5
          \not{\!\!D} \gamma^{\sigma} \not{\!\!D} \psi 
         -\overline{\psi} \gamma_5 D^\sigma\not{\!\!D}\psi
         -\overline{\psi} \gamma_5 \not{\!\!D} D^\sigma\psi ,\label{ops}\\
  E_2^\sigma &=& \overline{\psi} \gamma_5\partial^{\sigma}
          \not{\!\!D}\psi + \overline{\psi} \gamma_5 \not{\!\!D}
          \partial^{\sigma}\psi ,                         \nonumber\\
  E_3^\sigma &=& \overline{\psi} \gamma_5 \gamma^\sigma \not{\!\partial}
          \not{\!\!D}\psi + \overline{\psi} \gamma_5
          \not{\!\!D}\not{\!\partial}\gamma^\sigma\psi ,  \nonumber
\eea
where $D_\mu=\partial_\mu-igA_\mu^aT^a$ is the covariant derivative and
$\widetilde{G}_{\mu\nu}=\frac{1}{2}\varepsilon_{\mu\nu\alpha\beta}G^{\alpha\beta}$
is the dual field strength. For example, charge conjugation
(${\cal  C}$) and parity (${\cal P}$) transformations read as follows:
\bea
{\cal C}R_2^\sigma {\cal C}^{-1}&=&-g\psi^TC^{-1}
(-\widetilde{G}^{\sigma\mu})^T
\gamma_\mu C\overline{\psi}^T \nonumber\\
&=&-g\psi^T(\widetilde{G}^{\sigma\mu})^T\gamma_\mu^T\overline{\psi}^T\nonumber\\
&=&g\overline{\psi}\widetilde{G}^{\sigma\mu}\gamma_\mu\psi\nonumber\\
&=& R_2^\sigma,  \\
{\cal P}R_2^\sigma {\cal P}^{-1}&=&g\overline{\psi}\gamma_0
(-\widetilde{G}^\sigma_\mu)\gamma_\mu\gamma_0\psi\nonumber\\
&=&-g\overline{\psi}\widetilde{G}^{\sigma\mu}\gamma_\mu\psi\nonumber\\
&=& -R_2^\sigma. 
\eea

Here one should note that not all of the above operators are independent, as in
the case of twist-3 operators \cite{KYU}, and they are subject to the 
following constraint:
\be
R_1^{\sigma}=R_2^{\sigma}+E_1^{\sigma},
\ee
where we have used the identities, 
$D_\mu=\frac{1}{2}\{\gamma_\mu,\not{\!\!D}\}$ and
$[ D_\mu ,D_\nu ]=-igG_{\mu\nu}$.
Therefore any four operators out of (\ref{ops}) are independent and
they may mix through renormalization. 

Now, $E_1$ contains three gamma matrices, and when we rewrite the product
in terms of one gamma matrices, there occurs a mixing from the gauge
variant operator which is not an EOM operator.
Here we take $(R_2,E_1,E_2,E_3)$ to be the base of the independent operators.
The only operator which really contribute to the physical matrix element 
responsible for $\delta\Gamma_1$ is $R_2$.	
This twist-4 operator corresponds to the trace part of twist-3 operator,
$({\widehat
R}_2)_{\sigma\mu_1\mu_2}=g\overline{\psi}\widetilde{G}_{\sigma\{\mu_1}\gamma_{\mu_2\}}\psi -({\rm trace terms})$,
but there is no relation between the base for the twist-4 and that
for the twist-3 operators.

If we take this basis 
we have the following renormalization mixing matrix in the form of 
\be
\bmat{c} 
R_2 \\ E_1 \\ E_2 \\ E_3 
\emat_{\hspace{-0.1cm}R}
=
\bmat{cccc}
Z_{11} & Z_{12} & Z_{13} & Z_{14} \\
 0     & Z_{22} & Z_{23} & Z_{24} \\
0 & 0 & Z_{33} & 0 \\
0 & 0 & 0 & Z_{44} 
\emat
\bmat{c} 
R_2 \\ E_1 \\ E_2 \\ E_3 
\emat_{\hspace{-0.1cm}B}, 
\label{n3z1}
\ee
according to the general arguments in the following. 
(1) The counterterm for the EOM operators are given by the the EOM operators 
themselves. This is because the on-shell matrix
elements vanish for the EOM operators \cite{POLI}.
(2) A certain type of operators do not get renormalized.
And if we take those operators as one of the independent base,
the calculation becomes much simpler.
(3) The gauge variant operators also contribute to the mixing.

Now we turn to the calculation of the renormalization matrix
of this set of the operators.
Since the $R_2$ operator does not contribute to 2-point functions
but to 3-point functions with quarks $\psi$, $\overline{\psi}$ and
a gluon, $A$ as external lines, at the tree level, we only consider the 
following 3-point Green's function.
\bea
\Gamma_{O_\sigma}^{\psi\overline{\psi}A}\equiv \langle 0 |
T(\psi(p')A_{\rho}^a(l)\overline{\psi}(p)O_\sigma(0)) |0 \rangle^{1PI}, 
\eea
where these fields and the coupling constant involved are the bare quantities.
Here we employ the dimensional regularization and take the minimal
subtraction scheme. Note that we can not take all the external lines
on-shell except for a singular configuration of the momenta.
 
The Feynman rules are the following.
\bea
&&\left(\Gamma_{R_2^{\sigma}}^{\psi\overline{\psi}A}\right)_{\rm tree}=ig_B
\varepsilon_{\sigma\rho\alpha\beta} l^{\alpha}\gamma^{\beta}T^a ,\nonumber\\
&&\left(\Gamma_{E_1^{\sigma}}^{\psi\overline{\psi}A}\right)_{\rm tree}=
g_B\gamma_5\gamma_\sigma
(p+p')_\rho T^a -ig_B\varepsilon_{\sigma\rho\alpha\beta} 
l^{\alpha}\gamma^{\beta}T^a ,\nonumber\\
&&\left(\Gamma_{E_2^{\sigma}}^{\psi\overline{\psi}A}\right)_{\rm tree}=
-g_B\gamma_5\gamma_\rho (p+p')_\sigma T^a ,         \\ 
&&\left(\Gamma_{E_3^{\sigma}}^{\psi\overline{\psi}A}\right)_{\rm tree}=
g_B\gamma_5g_{\sigma\rho}
(\not{\! p}+\not{\! p'})T^a -g_B\gamma_5\gamma_\rho(p+p')_\sigma T^a \nonumber\\
&&\hspace{3cm}-g_B\gamma_5\gamma_\sigma(p+p')_\rho T^a+ig_B\varepsilon_{\sigma\rho\alpha\beta}
l^\alpha\gamma^\beta T^a.  \nonumber
\eea
The Feynman diagrams contributing to these Green's functions are in Fig.1.
The one-loop radiative corrections lead to
\bea
(\Gamma_{R_2}^{\psi\overline{\psi}A})_{\mbox{1-loop}}&=&
\left\{ 1+\pole [-\frac 3 4 C_2(G)]\right\}
ig_B\varepsilon_{\sigma\rho\alpha\beta} l^{\alpha}\gamma^{\beta}T^a\nonumber\\
&+& \pole [-\frac 3 5 C_2(R)+\frac 1 4 C_2(G)]
g_B\gamma_5\gamma_\sigma(p+p')_\rho T^a  \nonumber\\
&+& \pole [\frac 1 3 C_2(R)-\frac 1 4 C_2(G)]
g_B\gamma_5\gamma_\rho (p+p')_\sigma T^a  \label{r2loop}\\
&+& \pole \frac 1 3 C_2(R)
g_B\gamma_5g_{\sigma\rho}(\not{\! p}+\not{\! p'})T^a.\nonumber
\eea
It should be noted that the tensor structure  $ig_B\varepsilon_
{\sigma\rho\alpha\beta} l^{\alpha}\gamma^{\beta}T^a$  is 
also included in
$\left(\Gamma_{E_1^{\sigma}}^{\psi\overline{\psi}A}\right)_{\rm tree}$
and $\left(\Gamma_{E_3^{\sigma}}^{\psi\overline{\psi}A}\right)_{\rm
tree}$. So we can not connect directly the coefficient of the 
first line of (\ref{r2loop}) with the renormalization 
constant of $R_2$. We need to separate this coefficient into 
the $R_2$-part and the EOM-parts. In this sense EOM operators also contribute 
to the physical quantity $\gamma_{NS}$. 
Rearranging the results of (\ref{r2loop}) properly, we 
obtain; 
\bea
(\Gamma_{R_2}^{\psi\overline{\psi}A})_{\mbox{1-loop}}&=&
\left\{ 1+\pole [-\frac 5 3 C_2(R)+C_2(G)]\right\}
(\Gamma_{R_2}^{\psi\overline{\psi}A})_{\mbox{tree}} \nonumber\\
&+& \pole [-\frac 4 3 C_2(R)+\frac 1 4 C_2(G)]
(\Gamma_{E_1}^{\psi\overline{\psi}A})_{\mbox{tree}} \nonumber\\
&+& \pole [-\frac 2 3 C_2(R)+\frac 1 4 C_2(G)]
(\Gamma_{E_2}^{\psi\overline{\psi}A})_{\mbox{tree}} \\
&+& \pole \frac 1 3
C_2(R)(\Gamma_{E_3}^{\psi\overline{\psi}A})_{\mbox{tree}}.\nonumber
\eea

Now we introduce renormalization constants as follows:
\bea
A_B^\mu=Z_3^{1/2}A_R^\mu,\hspace{1cm}
\psi_B=Z_2^{1/2}\psi_R,\hspace{1cm}
\gb=Z_g\gr=Z_1Z_3^{-3/2}
\gr .
\eea
The composite operators are renormalized as
\be
(O_i)_R=\sum_j Z_{ij}(O_j)_B,
\ee
and the Green's functions of the composite operators with $\psi$, $\overline{\psi}$
and $A$ as the external lines are renormalized as follows:
\be
{\left(\Gamma_{O_i}\right)}_R =\sum_jZ_2\sqrt{Z_3}Z_{ij}
\left (\Gamma_{O_j}\right )_B.
\label{gam}
\ee
For example, $\left(\Gamma_{R_2}\right)_R$ reads
\bea
\left(\Gamma_{R_2}\right)_R
&=&Z_2Z_3^{1/2}\left[Z_{11}\left\{\left( 1+\alpole(-\frac 5 3 C_2(R)+C_2(G))
\right)\left(\Gamma_{R_2}\right)_{\mbox{tree}} \right.\right.\nonumber\\
&+&\alpole (-\frac 4 3 C_2(R)+\frac 1 4 C_2(G))
\left(\Gamma_{E_1}\right)_{\mbox{tree}} \nonumber\\
&+&\alpole (-\frac 2 3 C_2(R)+\frac 1 4 C_2(G))
\left(\Gamma_{E_2}\right)_{\mbox{tree}} \nonumber\\
&+&\left.\alpole\frac 1 3 C_2(R)\left(\Gamma_{E_3}\right)_{\mbox{tree}}\right\} \nonumber\\
&+&\alpole z_{12}\left(\Gamma_{E_1}\right)_{\mbox{tree}}
+\alpole z_{13}\left(\Gamma_{E_2}\right)_{\mbox{tree}}
+\left.\alpole z_{14}\left(\Gamma_{E_3}\right)_{\mbox{tree}}\right],
\label{gamr2r}
\eea
where $\widehat{\alpha}\equiv \displaystyle{\frac{g_R^2}{16\pi^2}}$.
Since we have in the Feynman gauge
\be
Z_2 Z_3^{1/2}=\left\{1-\alpole
[C_2(R)+C_2(G)]\right\}g_R g_B^{-1},
\ee
the above equation (\ref{gamr2r}) becomes
\bea
\left(\Gamma_{R_2}\right)_R
&=&1+\alpole [z_{11}-\frac 8 3 C_2(R)]
\left(\Gamma_{R_2}^{g_R^{}}\right)_{\mbox{tree}} \nonumber\\
&+&\alpole (z_{12}-\frac 4 3 C_2(R)+\frac 1 4 C_2(G))
\left(\Gamma_{E_1}^{g_R^{}}\right)_{\mbox{tree}} \nonumber\\
&+&\alpole (z_{13}-\frac 2 3 C_2(R)+\frac 1 4 C_2(G))
\left(\Gamma_{E_2}^{g_R^{}}\right)_{\mbox{tree}} \nonumber\\
&+&\alpole (z_{14}+\frac 1 3 C_2(R))
\left(\Gamma_{E_3}^{g_R^{}}\right)_{\mbox{tree}}, 
\label{gamr2e}
\eea
which should be finite. $z_{ij}$ is defined as, 
\be
Z_{ij}\equiv \delta_{ij}+\alpole z_{ij}, 
\ee
and $(\Gamma^{g_R}_{R_2})_{\rm tree}$ denotes  $(\Gamma_{R_2})_{\rm tree}$ 
with $g_B$ replaced by $g_R$. 
For $\Gamma_{E_i}^{\psi\overline{\psi}A}$, there are additional
diagrams due to tree 
quark-antiquark vertices (Fig.2). 
\bea
\left(\Gamma_{E_1}\right)_R
&=& Z_2Z_3^{1/2}\left[Z_{22}\left(\Gamma_{E_1}\right)_B+
Z_{23}\left(\Gamma_{E_2}\right)_B+Z_{24}\left(\Gamma_{E_3}\right)_B\right]
\nonumber\\
&=&1+\alpole [z_{22}- C_2(R)-\frac 1 2 C_2(G)]
\left(\Gamma_{E_1}^{g_R^{}}\right)_{\mbox{tree}} \nonumber\\
&+&\alpole [z_{23}+2C_2(R)+\frac 1 4 C_2(G)]
\left(\Gamma_{E_2}^{g_R^{}}\right)_{\mbox{tree}} \nonumber\\
&+&\alpole [z_{24}-C_2(R)-\frac 3 8 C_2(G)]
\left(\Gamma_{E_3}^{g_R^{}}\right)_{\mbox{tree}}. 
\label{game1r}
\eea
Further, the EOM operators like $E_2$ and $E_3$ which are of the form 
$E=\overline{\psi}B\displaystyle{\frac{\delta S}{\delta\overline{\psi}}}$, 
where $B$ is independent of fields, do not get renormalized \cite{COLL}. 
This can be seen as follows.
\bea
&&Z_2Z_3^{-1/2}
\langle 0 | T\left(\psi(x_1)A_{\rho}^a(x_2)\overline{\psi}(x_3)
\overline{\psi}B\displaystyle{\frac{\delta S}{\delta\overline{\psi}}}(y)
\right) |0 \rangle\nonumber\\
&&\hspace*{1.3cm}=-iZ_2Z_3^{-1/2}\int{\cal {D}}\psi
{\cal {D}}\overline{\psi}{\cal {D}}A \psi(x_1)A_{\rho}^a(x_2)
\overline{\psi}(x_3)\overline{\psi}B\displaystyle{\frac{\delta e^{iS}}
{\delta\overline{\psi}}}(y)  \nonumber\\
&&\hspace*{1.3cm}=-iZ_2Z_3^{-1/2}\int {\cal {D}}\psi{\cal {D}}\overline{\psi}{\cal {D}}A 
\left\{\psi(x_1)A_{\rho}^a(x_2)
\overline{\psi}(x_3) B e^{iS}\delta^4(0)\right.\\
&&\hspace*{6cm}\left. +\psi(x_1)A_{\rho}^a(x_2)\overline{\psi}(y) e^{iS}
\delta^4(y-x_3)\right\}
\nonumber\\
&&\hspace*{1.3cm}=-iZ_2Z_3^{-1/2}\langle 0 | 
T\left(\psi(x_1)A_{\rho}^a(x_2)\overline{\psi}B(y)\right) |0 \rangle
\delta^4(y-x_3)
\label{finite}\nonumber,
\eea
where the last equality holds as $\delta^4(0)$ vanishes in the sense 
of dimensional regularization. If $B$ does not contain any field, 
the RHS of (\ref{finite}) becomes finite by the wave functional
renormalization for $\psi, \overline{\psi}, A$. 
Therefore $E=\overline{\psi}B\displaystyle{\frac{\delta
S}{\delta\overline{\psi}}}$ is a finite operator.
Explicit calculations also indicate 
\be
\left(\Gamma_{E_2}\right)_R=Z_2Z_3^{1/2}Z_{33}\left(\Gamma_{E_2}\right)_B
=\left(1+\alpole z_{33}\right)\left(\Gamma_{E_2}^{g_R^{}}\right)_{\mbox{tree}},
\label{game2r}
\ee
and
\be
\left(\Gamma_{E_3}\right)_R=Z_2Z_3^{1/2}Z_{44}\left(\Gamma_{E_3}\right)_B
=\left(1+\alpole z_{44}\right)\left(\Gamma_{E_3}^{g_R^{}}\right)_{\mbox{tree}}.
\label{game3r}
\ee

From the finiteness of (\ref{gamr2r}), (\ref{game1r}), 
(\ref{game2r}), (\ref{game3r}), 
we get the following results for the renormalization constants:
\bea
\ba{ll}
z_{11}={8\over 3}C_2(R),  & z_{12}={4\over 3}C_2(R)-{1\over 4}C_2(G), \\ 
z_{13}={2\over 3}C_2(R)-{1\over 4}C_2(G), & z_{14}=-{1\over 3}C_2(R), \\
z_{22}=C_2(R)+{1\over 2}C_2(G),  & z_{23}=-2C_2(R)-{1\over 4}C_2(G), \\ 
z_{24}=C_2(R)+{3\over 8}C_2(G),  & z_{33}=z_{44}=0 .
\label{rm}
\ea
\eea
This result is in agreement with the general theorem on the renormalization
mixing matrix \cite{JL,COLL,H}. 
We should note that gauge variant EOM operator is  also necessary to 
renormalize the physical operator.

We now determine the anomalous dimension of $R_2^\sigma$ operator. 
In physical matrix elements, the EOM operators do not contribute 
and we have
\bea
\langle\mbox{phys}|(R_2^{\sigma})_B|\mbox{phys}\rangle &=&Z_{11}^{-1}
\langle\mbox{phys}|(R_2^{\sigma})_R|\mbox{phys}\rangle \nonumber\\
&=&\left[ 1-\frac{g^2}{16\pi^2}\frac{1}{\epsilon}\frac{8}{3}C_2(R)\right] 
\langle\mbox{phys}|(R_2^{\sigma})_R|\mbox{phys}\rangle.
\eea
Therefore the anomalous dimension $\gamma_{NS}$ turns out to be
\bea
\gamma_{NS}(g)&\equiv& Z_{11}\mu\frac{d}{d\mu}(Z_{11}^{-1}) \nonumber\\
&=&\frac{g^2}{16\pi^2}\cdot 2z_{11}+O(g^4) \nonumber\\
&=&\frac{g^2}{16\pi^2}\gamma_{NS}^0+O(g^4),
\eea
and
\be
\gamma_{NS}^0=2z_{11}=\frac{16}{3}C_2(R),
\ee
which coincides with the result obtained by Shuryak and Veinshtein \cite{SV}
using the background field method.

Including the twist-4 effect the Bjorken sum rule becomes 
\bea
&&\int_0^1 dx\left[g_1^p(x,Q^2)-g_1^n(x,Q^2) \right] \nonumber\\
&=&
\frac{1}{6}\left\{g_A\left(1-\frac{\alpha_s(Q^2)}{\pi}+{\cal O}(\alpha_s^2)
\right) 
-\frac{8}{9Q^2}f_3\left(\frac{\log{Q^2}/\log{\Lambda^2}}{\log{Q^2_0}/
\log{\Lambda^2}}\right)^{\displaystyle{-32/9\beta_0}}\right\}, 
\eea
in the case of QCD.

%section 4

\section{Flavor singlet part}
So far we have considered the flavor non-singlet part. Now we turn to the
flavor singlet component.
We should generally take account of gluon operators and BRS-exact 
operators as well in this case.\\
At first we see whether there exists other physical operators in
addition to $R_2$.
The possible twist-4 and spin-1 operators are the following:
\bea
&&\widetilde{G}^{\alpha}_{\mu}\hat{D}^{\sigma}G^{\mu}_{\alpha}
= \widetilde{G}^{\alpha}_{\mu}\hat{D}^{[\sigma}G^{\mu]}_{\alpha}
+\widetilde{G}^{\alpha}_{\mu}\hat{D}^{\mu}G^{\sigma}_{\alpha}
=0, \nonumber\\
&&\widetilde{G}^{\alpha}_{\mu}\hat{D}^{\mu}G^{\sigma}_{\alpha} 
= 0 \quad \mbox{(from Bianchi identity)}, \nonumber\\
&&\widetilde{G}^{\alpha\sigma}\hat{D}^{\mu}G_{\mu\alpha}\equiv R_3,
\eea
where $\hat{D}_\mu^{ab}\equiv \partial_\mu\delta^{ab} +f^{abc}T^c$.
Further, $R_3$, a gluon EOM operator and a BRS-exact 
operator may enter into the mixing. 
\bea
E_4&\equiv& \widetilde{G}^{\alpha\sigma}\hat{D}^\mu G_{\mu\alpha}
+\frac{1}{\alpha}\widetilde{G}^{\alpha\sigma}\partial_{\alpha}(\partial A)
-gf^{abc}\widetilde{G}^{\alpha\sigma}_a(\partial_\alpha\xi_b\omega_c)
-g\overline{\psi}\widetilde{G}^{\alpha\sigma}\gamma_\alpha\psi, \nonumber\\
A&\equiv& \frac{1}{\alpha}\widetilde{G}^{\alpha\sigma}\partial_{\alpha}(\partial A)
-gf^{abc}\widetilde{G}^{\alpha\sigma}_a(\partial_\alpha\xi_b\omega_c).
\eea
Then we have a trivial relation among them,  
\be
R_3=E_4+R_2-A.
\ee
Therefore it happens that the physical operator is still only $R_2$ 
even in the flavor singlet case. This is characteristic to the 
lowest-spin case of higher-twist, in which only a few tensor
structures are possible. 

It should be noted that  
we have only to consider the mixing between $R_2^{\sigma}$ 
and
\be
E_G^{\sigma}=\widetilde{G}^{\alpha\sigma}\hat{D}^{\mu}G_{\mu\alpha}-
g\overline{\psi}\gamma_{\alpha}\widetilde{G}^{\alpha\sigma}\psi, 
\ee
as long as we want to obtain the physical matrix element $Z_{11}^S$
where we do not have to consider Green's functions with ghost operators 
in their external legs \cite{GW}.\\
The mixing between $R_2$ and $E_G$ can be studied by computing
the Green's functions with two-gluon external lines shown in Fig.3.
Then we have
\be
\Gamma_{R_2}^{AA}(q)=\pole (-\frac 2 3 N_f)\times 
\left(\Gamma_{E_G}^{AA}\right)_{\mbox{tree}},
\ee
where $N_f$ denotes the number of fermions. 
$\left(\Gamma_{E_G}^{AA}\right)_{\mbox{tree}}$ is given by
\be
\left(\Gamma_{E_G}^{AA}\right)_{\mbox{tree}}=
i\varepsilon^{\sigma\rho\alpha\beta}q_{\alpha}(q^2g_{\beta\lambda}
-q_{\beta}q_{\lambda})
-i\varepsilon^{\sigma\lambda\alpha\beta}q_{\alpha}(q^2g_{\beta\rho}
-q_{\beta}q_{\rho}).
\ee
Now we introduce the renormalization constant $Z_{15}$ as 
\be
\left(R_2\right)_R=Z_{11}\left(R_2\right)_B+Z_{12}\left(E_1\right)_B
+Z_{13}\left(E_2\right)_B+Z_{14}\left(E_3\right)_B
+Z_{15}\left(E_G\right)_B.
\ee
So we get
\bea
\left(\Gamma_{R_2}^{AA}\right)_R&=&
Z_3\left[Z_{11}\left(\Gamma_{R_2}^{AA}\right)_B+
\sum_{i=1}^{3}Z_{1i}\left(\Gamma_{E_i}^{AA}\right)_B
+Z_{15}\left(\Gamma_{E_G}^{AA}\right)_B\right] \nonumber\\
&\sim& \alpole(-\frac 2 3 N_f)\left(\Gamma_{E_G}^{AA}\right)_{\mbox{tree}}
+Z_{15}\left(\Gamma_{E_G}^{AA}\right)_{\mbox{tree}}.
\eea
From the finiteness of the above expression, we have
\be
Z_{15}=\alpole\times\frac 2 3 N_f.
\ee

On the other hand, the relation (\ref{gam}) becomes 
\be
\left(R_2^{\psi\overline{\psi}A}\right)_R=Z_2Z_3^{1/2}
\left[Z_{11}\left(\Gamma_{R_2}^{\psi\overline{\psi}A}\right)_B+
\sum_{i=1}^3Z_{1i}\left(\Gamma_{E_i}^{\psi\overline{\psi}A}\right)_B
+Z_{15}\left(\Gamma_{E_G}^{\psi\overline{\psi}A}\right)_B\right],
\ee
in this case.
Since $(\Gamma^{\psi\overline{\psi}A}_{R_2})_B$ is just the same as the
flavor non-singlet case, we can easily extract $Z_{11}^S$ from the 
result of $Z_{11}^{NS}$.
\be
Z_{11}^{S}=Z_{11}^{NS}+\frac 2 3 N_f, 
\ee
namely,
\be
\gamma_S^0=\gamma_{NS}^0+\frac 4 3 N_f, 
\ee
hence we obtain the exponent for the singlet part
\be
-\frac{\gamma_S^0}{2\beta_0}=-\frac{\gamma_{NS}^0}{2\beta_0}
-\frac{2}{3}\frac{N_f}{\beta_0}
=-\frac {1}{\beta_0}\left(\frac{32}{9} + \frac{2}{3}N_f\right).
\ee
Again we reproduce the result of \cite{SV}. 
Substituting these results into (\ref{tw4}), we get   
\bea
\delta\Gamma_{\rm tw-4}^{\rm singlet}=\frac{C^{\rm twist-4}}{Q^2}\left(
\frac{\log{Q^2}/\log{\Lambda^2}}{\log{Q_0^2}/\log{\Lambda^2}}
\right)^{-(32/9\beta_0+2 N_f/3)}, 
\eea
where $C^{\rm twist-4}$ denotes the coefficients including the reduced 
matrix elements, $f_i$. 

\section{Concluding remarks}

In this paper, we have investigated the renormalization of the twist-4
operators relevant for the Bjorken and Ellis-Jaffe sum rules. We obtained 
the anomalous dimensions by considering off-shell Green's 
functions in which the role of EOM
operators was important. 
We have also seen that the renormalization
matrix satisfies
the general theorems for the renormalization of composite operators in
the gauge theory. Since 
there appears only one physical operator, our case seems to be 
the simplest example which contains non-trivial structures.  

Phenomenologically whether the twist-4 effects has significant 
contributions depends on the value of their reduced matrix elements.
Estimations of their values were given by various methods. For
example, from the QCD sum rules \cite{QCD}, $\delta\Gamma_1^{p-n}$ 
amounts to about 4{\%} of $\Gamma_1^{p-n}$, and the chiral-bag model \cite{JU} 
predicts similar value, but some other estimations predict twice or 
much larger values \cite{DIQ}. Furthermore the reduced matrix elements
of twist-4 operators are considered to be related to the renormalon ambiguity
of the coefficient functions of the leading twist parts \cite{Renom}.  
So we can not give any definite predictions about the reduced matrix
elements until some consistent treatments are developed for these problems.
On the other hand, the anomalous dimensions of twist-4 operators are 
free from the renormalon ambiguity and have definite values. 

The numerical values of the exponents of log-terms in
$\delta\Gamma_1^{p,n}$ are so small 
that the $\log{Q^2}$ scalings are quite hard to observe in experiments.
Since the anomalous dimensions are considered to become large 
as the spin of the operators increases,
the scaling of the twist-4 terms are probably easier to measure 
in some higher moments .  
However to obtain the anomalous dimensions in spin-n case 
by our method is not so easy because many gauge
variant EOM operators may enter into the operator mixing. 
In the twist-4 case, it is hard to simplify these structures 
by the method used in the twist-3 case \cite{KT,YY} in which 
many tree amplitude of different EOM operators are projected out to 
a single one by a null vector $\Omega_\rho$. Since the twist-4 
operators have a pair of Lorentz indices contracted, 
there still 
remain so many independent tree amplitudes after the contraction  
by $\Omega_\rho$ and this complicates  
the situation. So we need some further developments in the 
techniques to disentangle the mixing problem.

\vspace{1.0cm}
We are very grateful to J.Kodaira, T.Uematsu and Y.Yasui for valuable 
discussions. Thanks are also due to T.Uematsu for careful reading 
of the manuscript. 
%%%%%%%%%%%%%%%%%%%%%%%%%%%%%%%%%%%%%%%%%%%%%%
\newpage

\newpage
\input epsf.sty
\begin{figure}
\centerline{
\epsfxsize=14cm
\epsfbox{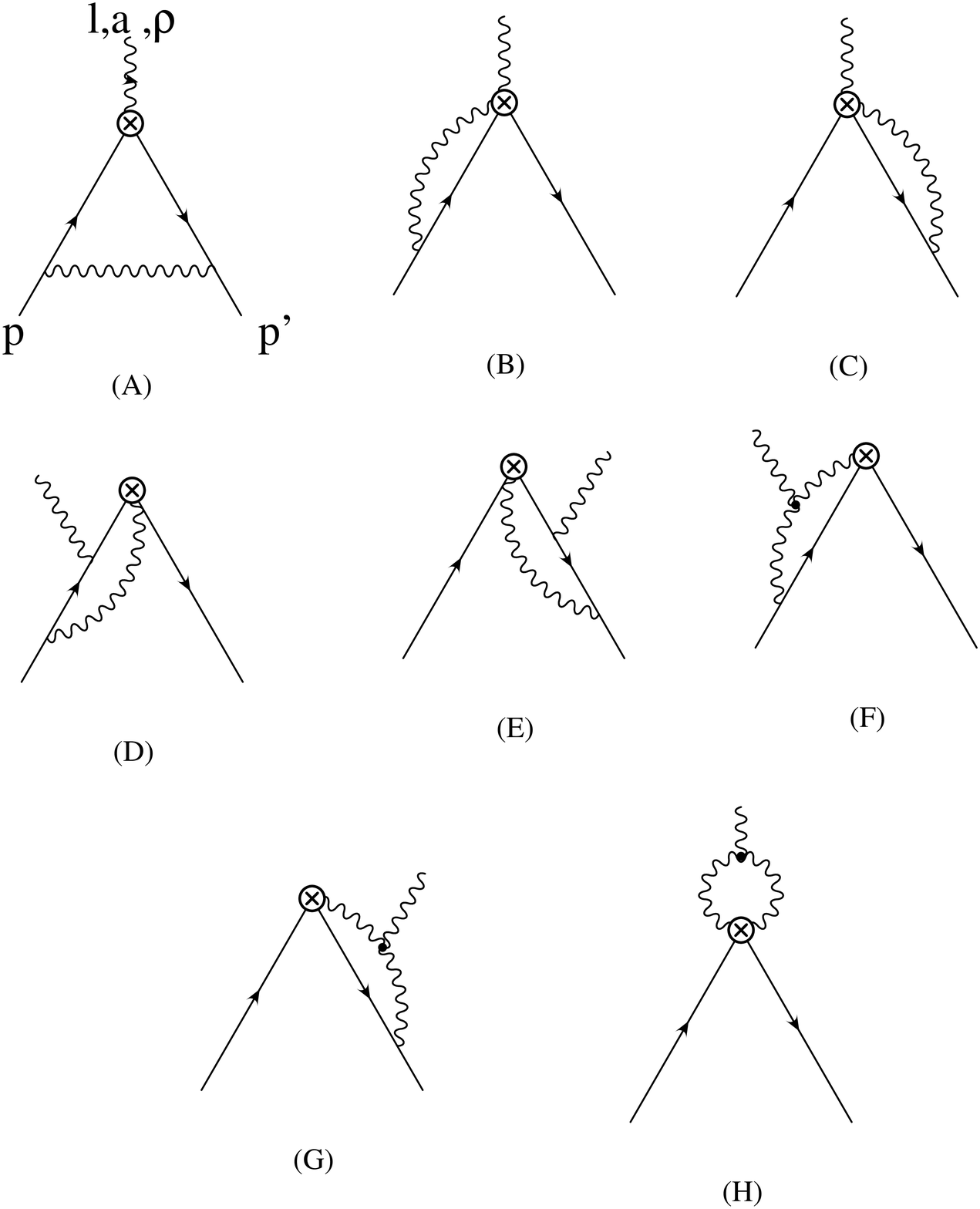}}
\caption{Feynman diagrams for $\Gamma_{R_2}^{\psi\overline{\psi}A}$}
\end{figure}

\newpage
\input epsf.sty
\begin{figure}
\centerline{
\epsfxsize=14cm
\epsfbox{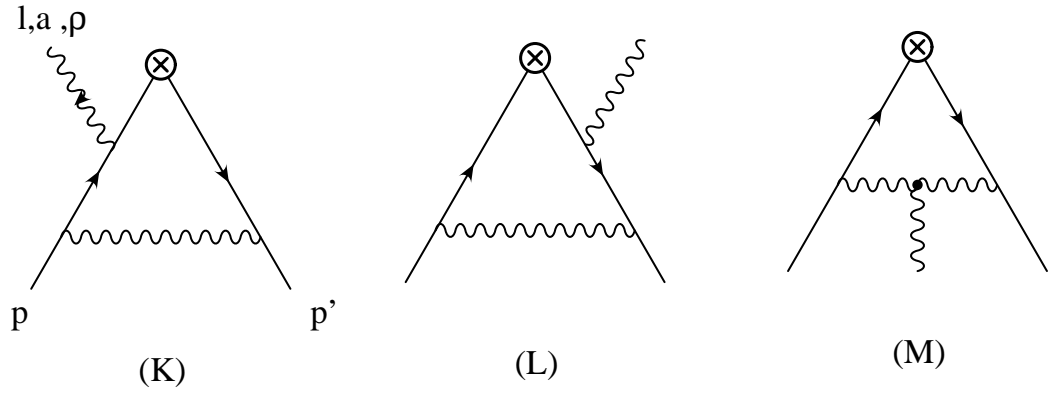}}
\caption{Additional diagrams for $\Gamma_{E_i}^{\psi\overline{\psi}A}$}
\end{figure}

\input epsf.sty
\begin{figure}
\centerline{
\epsfxsize=11cm
\epsfbox{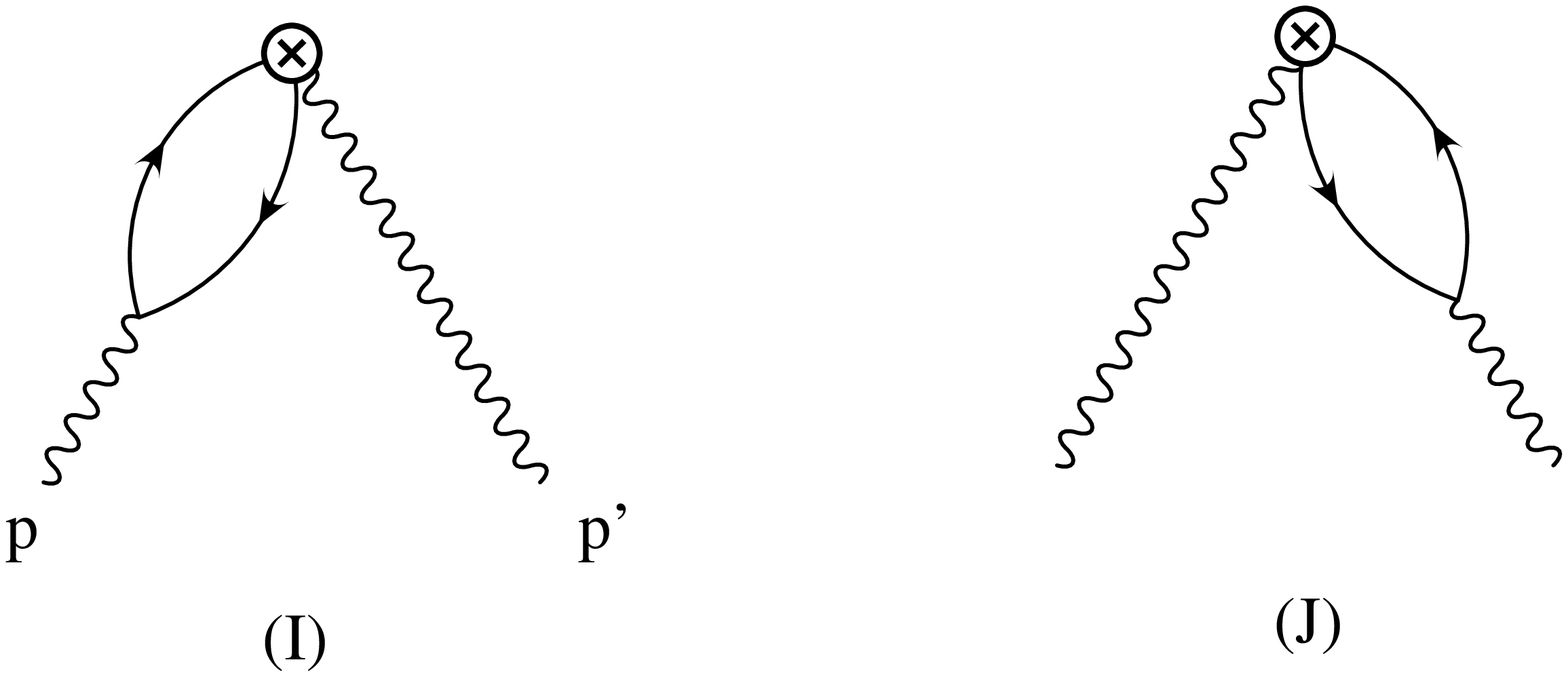}}
\caption{Feynman diagrams for $\Gamma_{R_2}^{AA}$}
\end{figure}


\begin{thebibliography}{99}

\bibitem{EMC}
     J.~Ashman {\it et al.} {\sl Phys.Lett.} {\bf B206} (1988) 364;\\
     V.~W.~Hughes {\it et al.} {\sl Phys.Lett.} {\bf B212} (1988) 511.

\bibitem{SMC}
     B.~Adeva {\it et al.} {\sl Phys.Lett.} {\bf B302} (1993) 553;
     {\bf B320} (1994) 400 \\
     D.~Adams {\it et al.}, {\sl Phys.Lett.} {\bf B329} (1994) 399;
     {\bf B336} (1994) 125. 

\bibitem{E142}
     P.~L.~Anthony {\it et al.} {\sl Phys.Rev.Lett.} {\bf 71} (1993) 959.

\bibitem{E143}
     K.~Abe {\it et al.} {\sl Phys.Rev.Lett.} {\bf 74} (1995) 346;
     {\sl Phys.Rev.Lett.} {\bf 75} (1995) 25;
     Preprint SLAC-PUB-95-6982 (hep-ex/9511013).

\bibitem{REV}
    For the recent theoretical review see for example, \\
    J. Ellis and M. Karliner, CERN-TH/95-279 (hep-ph/9510402);\\
    S. Forte, CERN-TH/95-305(hep-ph/9511345);\\
    B. L. Ioffe, ITEP No.62-95 (hep-ph/9511401).


\bibitem{EJ}
J.~Ellis and R.~L.~Jaffe, {\sl Phys.~Rev.}~{\bf D9} (1974) 1444;{\bf D10} 
(1974) 1669.

\bibitem{Bj}
 J.~D.~Bjorken, {\sl Phys.~Rev.}~{\bf 148} (1966) 1467;{\bf D1} (1970) 1376.

\bibitem{JL}
S.~D.~Joglekar and B.~W.~Lee, {\sl Ann.Phys.}(N.Y.){\bf 97}(1976)160.

\bibitem{COLL}
     J.~C.~Collins, {\it Renormalization} (Cambridge Univ. Press, 1984);
     and {\it references therein}.

\bibitem{SV}
    E.~V.~Shuryak and A.~I.~Vainshtein, {\sl Nucl.Phys.}
                    {\bf B201} (1982) 141;\\
    A.~P.~Bukhvostov, E.~A.~Kuraev and L.~N.~Lipatov, {\sl JETP Letters}
                    {\bf 37} (1984) 483.

\bibitem{KOD}
     J.~Kodaira, S.~Matsuda, K.~Sasaki and T.~Uematsu, {\sl Nucl.Phys.}
                                ~{\bf B159} (1979) 99;\\
     J.~Kodaira, S.~Matsuda, T.~Muta, K.~Sasaki and T.~Uematsu,
                    {\sl Phys.Rev.} {\bf D20} (1979) 627.

\bibitem{K}
     J.~Kodaira, {\sl Nucl.~Phys.}~{\bf B165} (1980) 129.

\bibitem{GL}
S.~G.~Gorishny and S.~A.~Larin, {\sl Phys.~Lett.}~{\bf B172}~(1986)~109;\\
S.~A.~Larin and J.~A.~M.~Vermaseren, {\sl Phys.~Lett.}~{\bf B259}~(1991)~345;\\
S.~A.~Larin, F.~V.~Tkachev and J.~A.~M.~Vermaseren, {\sl Phys. Rev. Lett.}
{\bf 66}~(1991)~862;\\
A.~L.~Kataev and V.~Starshenko, preprint CERN-TH-7198-94.

\bibitem{HKU}
 H.~Kawamura and T.~Uematsu, {\sl Phys. Lett.}~{\bf B343}~(1995)~346.

\bibitem{H}
     M.~Henneaux, {\sl Phys.~Lett.}~{\bf B313}(1993)35.

\bibitem{KYU}
     J.~Kodaira, Y.~Yasui and T.~Uematsu,
     {\sl Phys. Lett.}~{B344}~(1995)~348.

\bibitem{POLI}
     H.~D.~Politzer, {\sl Nucl.Phys.} {\bf B172} (1980) 349.
\bibitem{GW}
D.~J.~Gross and F.~Wilczek, {\sl Phys.~Rev.}~{\bf D9} (1974) 980.

\bibitem{QCD}
I.~I.~Balitsky and V.M.Braun, {\sl Nucl.Phys.} {\bf B311}(1989) 541; \\
G.~G.~Ross and R.~G.~Roberts, {\sl Phys.Lett.} {\bf B322}(1994) 425; \\
L.~Mankiewicz, E.~Stein and A.~Sh{\"a}fer, hep-ph 9510418.

\bibitem{JU}
X.~Ji and P.~Unrau, {\sl Phys.~Lett.} {\bf B333} (1994) 228.

\bibitem{DIQ}
M.~Anselmino, F.~Caruso and E.~Levin, {\sl Phys.Lett} {\bf B358}(1995)109;\\
B.~L.~Ioffe, hep-ph 9511264

\bibitem{Renom}
V.~M.~Braun, hep-ph 9505317;
A.~H.~Mueller, {\sl Phys.Lett.} {\bf B308}(1993)355;\\
M.~Meyer-Hermann, M.~Maul, L.~Mankiewicz, E.~Stein, A.~Sch{\"a}fer,\\ 
hep-ph 9605229.


\bibitem{KT}
Y.~Koike and K.~Tanaka, {\sl Phys.Rev.} {\bf D51} (1995) 6125.

\bibitem{YY}
J.~Kodaira, Y.~Yasui, K.~Tanaka and T.~Uematsu, hep-ph 9603377.  
\end{thebibliography}
\end{document}